\documentclass[preprint]{elsarticle}

\usepackage{lineno}
\modulolinenumbers[1]
\usepackage{hyperref} 

\usepackage{soul}

\usepackage{lipsum}
\usepackage{pstricks, pst-node, psfrag}
\usepackage{amssymb,amsmath}
\usepackage{mathtools}
\usepackage{verbatim,enumerate}
\usepackage{rotating, lscape}
\usepackage{setspace}
\usepackage[hang, flushmargin]{footmisc}
\usepackage{subfig}
\usepackage{caption}
\usepackage{cancel}
\usepackage{float}
\usepackage{ragged2e}
\usepackage{centernot}
\usepackage{tikz}
\usepackage{tikzsymbols}
\usetikzlibrary{shapes,arrows,positioning,plotmarks}
\usepackage{textcomp} 
\usepackage{gensymb} 
\usepackage{multicol}
\usepackage[boxruled, linesnumbered]{algorithm2e}
\SetKwComment{Comment}{$\triangleright$\ }{}
\usepackage{enumitem}
\setenumerate{noitemsep}

\usepackage{titlesec}
\titleformat{\paragraph}
{\normalfont\normalsize\bfseries}{\theparagraph}{1em}{}
\titlespacing*{\paragraph}
{0pt}{3.25ex plus 1ex minus 0.2ex}{1.5ex plus 0.2ex}

\tikzstyle{endpt} = [rectangle, draw, fill=red!20,
    text width=9.8em, text centered, rounded corners, minimum height=4em]
\tikzstyle{block} = [rectangle, draw, top color=white, bottom color=blue!20,
    text width=9.8em, text centered, rounded corners, minimum height=4em]
\tikzstyle{line} = [draw, -latex', very thick]

\usepackage{listings}
\usepackage{color} 
\definecolor{mygreen}{rgb}{28,172,0} 
\definecolor{mylilas}{rgb}{170,55,241}
\definecolor{mygray}{rgb}{0.5,0.5,0.5}
\definecolor{mycyan}{rgb}{0,255,255}
\definecolor{magenta}{rgb}{1,0,1}

\definecolor{backgreen}{rgb}{0.00, 0.169, 0.212}
\definecolor{textgray}{rgb}{0.514, 0.580, 0.589}

\usepackage[T1]{fontenc}
\newcommand*\patchAmsMathEnvironmentForLineno[1]{%
    \expandafter\let\csname old#1\expandafter\endcsname\csname #1\endcsname
    \expandafter\let\csname oldend#1\expandafter\endcsname\csname end#1\endcsname
    \renewenvironment{#1}%
        {\linenomath\csname old#1\endcsname}%
        {\csname oldend#1\endcsname\endlinenomath}}%
\newcommand*\patchBothAmsMathEnvironmentsForLineno[1]{%
    \patchAmsMathEnvironmentForLineno{#1}%
    \patchAmsMathEnvironmentForLineno{#1*}}%
\AtBeginDocument{%
    \patchBothAmsMathEnvironmentsForLineno{equation}%
    \patchBothAmsMathEnvironmentsForLineno{align}%
    \patchBothAmsMathEnvironmentsForLineno{flalign}%
    \patchBothAmsMathEnvironmentsForLineno{alignat}%
    \patchBothAmsMathEnvironmentsForLineno{gather}%
    \patchBothAmsMathEnvironmentsForLineno{multline}%
}

\newcommand{\R}{\mathbb{R}}

\newcommand{\cD}{\mathcal{D}}

\newcommand{\cN}{\mathcal{N}}
\newcommand{\cO}{\mathcal{O}}

\newcommand{\cW}{\mathcal{W}}
\newcommand{\cX}{\mathcal{X}}

\newcommand{\defeq}{\coloneqq}

\newcommand{\pder}[2][]{\frac{\partial#1}{\partial#2}}

\newcommand{\dt}{\Delta t}

\renewcommand{\vec}[1]{\boldsymbol{#1}}

\renewcommand{\tilde}{\widetilde}
\renewcommand{\hat}{\widehat}

\DeclareMathOperator{\erf}{erf}

\DeclareMathOperator{\diag}{diag}

\newcommand{\rev}[1]{\textcolor{black}{#1}}

\journal{Advances in Water Resources}

\biboptions{sort&compress}
\begin{document}

\begin{frontmatter}

\title{A Mass-transfer Particle-tracking Method for Simulating Transport with Discontinuous Diffusion Coefficients\tnoteref{mytitlenote}}
\tnotetext[mytitlenote]{This work was partially supported by the US Army Research Office under Contract/Grant number W911NF-18-1-0338; the National Science Foundation under awards EAR-1417145 and DMS-1614586; and the DOE Office of Science under award DE-SC0019123.}

\author{Michael J. Schmidt\fnref{nd}}
\ead{mschmi23@nd.edu}
\author{Nicholas B. Engdahl\fnref{wsu}}
\ead{nick.engdahl@wsu.edu}
\author{Stephen D. Pankavich\fnref{ams}}
\ead{pankavic@mines.edu}
\author{Diogo Bolster\fnref{nd}}
\ead{bolster@nd.edu}

\fntext[nd]{Department of Civil and Environmental Engineering and Earth Sciences, University of Notre Dame, Notre Dame, IN, 46556, USA}
\fntext[ams]{Department of Applied Mathematics and Statistics, Colorado School of Mines, Golden, CO, 80401, USA}
\fntext[wsu]{Department of Civil and Environmental Engineering, Washington State University, Pullman, WA, 99164, USA}

\begin{abstract}

The problem of a spatially discontinuous diffusion coefficient ($D(\vec x)$) is one that may be encountered in hydrogeologic systems due to natural geological features or as a consequence of numerical discretization of flow properties.
To date, mass-transfer particle-tracking (MTPT) methods, a family of Lagrangian methods in which diffusion is jointly simulated by random walk and diffusive mass transfers, have been unable to solve this problem.
This manuscript presents a new mass-transfer (MT) algorithm that enables MTPT methods to accurately solve the problem of discontinuous $D(\vec x)$.
To achieve this, we derive a semi-analytical solution to the discontinuous $D(\vec x)$ problem by employing a predictor-corrector approach, and we use this semi-analytical solution as the weighting function in a reformulated MT algorithm.
This semi-analytical solution is generalized for cases with multiple 1D interfaces as well as for 2D cases, including a $2 \times 2$ tiling of 4 subdomains that corresponds to a numerically-generated diffusion field.
The solutions generated by this new mass-transfer algorithm closely agree with an analytical 1D solution or, in more complicated cases, trusted numerical results, demonstrating the success of our proposed approach.

\end{abstract}

\begin{keyword}
Lagrangian Modeling
\sep
Particle Methods
\sep
Mass-transfer particle-tracking
\sep
Imperfect Mixing
\sep
Diffusion-reaction Equation
\sep
Composite Porous Media
\sep
Discontinuous Diffusion Coefficients
\end{keyword}

\end{frontmatter}


\section{Introduction} 
\label{sec:intro}

Simulating diffusive transport under the condition of a spatially discontinuous diffusion coefficient is a challenging problem that is frequently encountered in hydrogeological contexts \cite{uffink1983,labolle_composite,appuhamillage_trans_interface,semra_reflect,oukili_negMass_discontD}.
Physically, this can occur wherever there is an abrupt  change in the material properties of a medium, like the sharp interfaces between different depositional units.
Sharp discontinuities can also be seen in, for example: fractured or composite media, local compaction zones, or at the interface between a saturated and unsaturated zone.
From a numerical perspective, any non-constant hydraulic conductivity field that is discretized will generate a diffusion/dispersion field containing numerous discontinuities.
Interpolation or averaging methods have been used in the past to smooth these discontinuities, and these can be effective as long as the differences in magnitude of the parameter(s) across the interface is \rev{sufficiently small (in general, less than an order of magnitude)}.
However, when the difference in diffusion coefficients between cells, or regions of a domain, becomes \rev{sufficiently large}, the simplest versions of these methods can fail, and overcoming this challenge requires a more nuanced approach.

Random-walk particle-tracking (RWPT) methods are a class of stochastic Lagrangian (mesh-free or gridless) methods that are commonly used to simulate advective-diffusive transport.
These methods were originally formulated in the context of conservative (non-chemically reactive) transport or cases of simple, linear reactions, such as sorbing solutes or first-order decay \cite{labolle_1996,Salamon_2006}.
They are popular because they introduce no numerical diffusion into the simulation of the advection (hyperbolic) operator, and they also escape the burden imposed by restrictive stability conditions in Eulerian (grid-based) methods, resulting in lower run times than corresponding Eulerian methods \cite{Benson_AWR_2016}.
Further, because RWPT is a stochastic algorithm, statistics of concentrations can be readily generated instead of expected values (point estimates).
In this context, the problem of discontinuous diffusion coefficients has received much attention, resulting in various methods for overcoming the difficulties of simulating such a system \cite[e.g.,][]{uffink1983,appuhamillage_trans_interface,semra_reflect,Hoteit2002,bechtold_reflect_nonlinDT,oukili_negMass_discontD,bagtzoglou_discontD_interp,labolle_1996,labolle_composite}, each with their own merits and drawbacks.

One of the major advantages of classical RWPT is its speed, due to the fact that every particle is completely independent of its neighbors.
However, this also means that complex reactions cannot be simulated since particle interactions are not allowed.
Recent developments in the field of RWPT have enabled methods to simulate complex and nonlinear chemical reactions in the presence of transport using either collision-based reactions between particles of opposite species \cite{Benson_react,Paster_JCP,Bolster_mass,schmidt_kRPT,guillem2017kde,guillem_adaptive2018}, or by treating individual particles as reaction volumes that communicate via diffusive mass transfers \cite{Benson_arbitrary}.
The latter, referred to as mass-transfer particle-tracking (MTPT) algorithms, offer the increased flexibility of being able to model arbitrarily complex chemical reactions at relatively low computational cost \cite{Engdahl_WRR,schmidt_hMetal}, including generalized ``reactions'' such as the aging of water parcels \cite{benson_aging19}.
The mass-transfer (MT) portion of these MTPT methods has been demonstrated to solve the diffusion equation to $\cO(\dt)$ \cite{mass_trans_acc} and exhibit superlinear convergence as particle numbers grow large \cite{Schmidt_fluid_solid}.
Additionally, a method for parallelizing the MTPT method via domain decomposition has recently been developed and achieves linear speedup up to hundreds of computational cores/subdomains \cite{engdahl_ddc}.
MTPT methods have also been shown to be related to smoothed-particle hydrodynamics (SPH) methods \cite[e.g.,][]{herrera_2009,herrera_2013,Gingold_originalSPH,Monaghan_SPHappl} under specific modeling choices, including the use of a Gaussian spatial kernel \cite{guillem_SPH_equiv}.
Despite these advances, past work on MTPT methods has neither addressed the impact of a discrete parameter field on the mass transfer operations nor accounted for the possible errors that may be incurred.

All previously-mentioned random-walk methods may be employed for diffusion coefficients with spatial discontinuities because they are capable of simulating small-scale mixing and non-mixed spreading of solute separately \cite{benson_mix_spread}.
In other words, spreading may be simulated by a random walk and mixing as a mass transfer.
However, accuracy of the mass-transfer step is only preserved for a smoothly varying field (i.e., one in which interpolation may be reasonably performed), and the current MTPT schemes incur error when there is a sharp discontinuity.
This is similar to the problems identified by \cite{labolle_composite} for classical RWPT.
MTPT has clear applications for highly accurate simulations of mixing-limited reactive transport, but this issue undermines its accuracy.
Thus, the purpose of the current paper is to address this deficiency and ensure that MTPT methods remain accurate even in such a case.

In Section \ref{sec:model}, we outline the specific mathematical problem on which we will focus, and introduce the methods used to solve the problem in Section \ref{sec:computational_methods}.
In Section \ref{sub:RWPT_method}, we provide a brief overview of RWPT methods and discuss how the problem of discontinuous diffusion coefficients is typically handled, with specific focus on a particular predictor-corrector technique \cite{labolle_composite} that we extend to MTPT.
In Section \ref{sub:MTPT_method}, we outline our approach to solving the discontinuous diffusion coefficient problem with an MTPT method by employing an alternative mass-transfer kernel.
Section \ref{sec:results} is devoted to discussing the results of applying the new MTPT method.
Finally, Section \ref{sec:conclusions} presents the conclusions drawn from our work.


\section{Analytic model} 
\label{sec:model}

We consider a chemically-conservative, single species, purely diffusive system that may be described by the (heterogeneous) diffusion equation
\begin{equation}\label{diffEqn}
    \pder[C]{t} = \nabla \cdot  \left(D (\vec x) \nabla C\right),\qquad \vec x \in \Omega \subseteq \R^d,\qquad t > 0,
\end{equation}
where $C(t, \vec x)$ $[\text{mol}\ \text{L}^{-d}]$ is the concentration of the single species, $D(\vec x)$ $[\text{L}^2\ \text{T}^{-1}]$ is the scalar diffusion coefficient, which, for our purposes, may be a function of space.
For this work, we concern ourselves with the condition where $D$ may be discontinuous.
This case leads to infinite spatial derivatives at all discontinuities, so the question is how best to numerically evaluate the $\nabla \cdot D$ term within the chosen method to minimize artifacts of the discontinuity.
The discontinuity we use for this study is created by partitioning the domain, $\Omega$, into $N_\Omega$ subdomains such that $\Omega = \Omega_1 \cup \Omega_2 \cup \dots \cup \Omega_{N_\Omega}$, where each subdomain has its own constant-valued diffusion coefficient, $D_i,\ i = 1, \dots, N_\Omega$, and the interface between subdomains $i$ and $j$ is denoted $\gamma_{ij}$.


\section{Computational methods} 
\label{sec:computational_methods}

\subsection{Random-walk particle-tracking Method} 
\label{sub:RWPT_method}

The classical Lagrangian method for simulating the system of interest is standard random-walk particle-tracking (RWPT)  \cite{Thompson_1987,labolle_1996}.
In these methods, masses are divided among particles that simulate diffusion via the Langevin equation (formulated for homogeneous $D$)
\begin{equation}
    \vec X_i(t + \dt) = \vec X_i(t) + \vec \xi_i \sqrt{2 D \dt},
\end{equation}
where $\vec X_i(t)$ is the position of particle $i$ at time $t$, $\dt$ is the chosen simulated  timestep, and $\vec \xi_i$ is a $d$-dimensional vector of random numbers drawn from a standard normal, $\cN(0, 1)$, distribution.
In this basic form, RWPT methods are unable to simulate the problem of discontinuous diffusion coefficients ($D(\vec x)$), described in Section \ref{sec:model}.
Conceptually, the problem is that, during the course of a single-step random walk, a particle may ``see'' diffusion at the rates on both sides of the discontinuity in $D(\vec x)$; however, there are well-documented strategies for overcoming this.

The first general group of strategies are reflection methods \cite{uffink1983,appuhamillage_trans_interface,semra_reflect,Hoteit2002}, which may include a nonlinearly decomposed time step \cite{bechtold_reflect_nonlinDT}, interpolation methods \cite{bagtzoglou_discontD_interp}, or a combination thereof \cite{labolle_1996}.
A selection of these are reviewed and compared in \cite{labolle_1998}\rev{, and the conclusion reached therein is that, among the methods considered, those of \cite{uffink1983,semra_reflect} provide the best accuracy.}
\rev{A benchmark comparison of various methods is also conducted by \cite{lejay_benchmark} who distinguish between methods that preserve or lose important physical and numerical properties, and recent work of the same authors presents a method that employs skew Brownian motion densities with exponential timestepping to capture the dynamics of the discontinuous $D(\vec x)$ problem \cite{lejay_disco_expDT}.}
Another recent approach \cite{oukili_negMass_discontD} employs negative-mass particles in a partial reflection scheme, so as to keep the total mass in a system constant and maintain particle independence.

To demonstrate how discontinuous $D(\vec x)$ is handled with RWPT, and because we later use this method to generate reference solutions, we briefly discuss the work of \emph{Labolle et al.} \cite{labolle_composite}.
We consider this method because it bears resemblance to the algorithm we present in Section \ref{sub:MTPT_method}.
Also, it is relatively simple to implement, and the extension to greater than one spatial dimension is straightforward, unlike some other approaches.
This method may be thought of as a predictor-corrector approach, and is formulated as
\begin{align}
    \vec \cX_i &= \vec X_i(t) + \vec \xi_i \sqrt{2 D(\vec X_i) \dt}. \label{labolle_predict}\\
    \vec X_i(t + \dt) &= \vec X_i(t) + \vec \xi_i \sqrt{2 D(\vec \cX_i) \dt},\label{labolle_correct}
\end{align}
In words, a ``predictor'' random walk is first taken from $\vec X_i(t)$ to $\vec \cX_i$ in \eqref{labolle_predict} to determine the diffusion coefficient that is then used in the ``corrected'' random walk from $\vec X_i(t)$ to $\vec X_i(t + \dt)$ in \eqref{labolle_correct}.
A subtle but important point is that the same random number, $\vec \xi_i$ must be used in \eqref{labolle_correct} that was generated for \eqref{labolle_predict}.


\subsection{Mass-transfer particle-tracking method} 
\label{sub:MTPT_method}

Another family of Lagrangian methods that has gained attention recently are the mass-transfer particle-tracking (MTPT) methods, which are the focus of this work. \cite{Benson_arbitrary,Engdahl_WRR,mass_trans_acc,Schmidt_fluid_solid,schmidt_hMetal}.
These methods are quite similar to RWPT methods in that diffusion is typically still simulated, in part, by random walks.
However, the important distinction is that particle masses are no longer fixed and can be transferred among particles according to an algorithm that also simulates diffusion.
The MT algorithm may be given as
\begin{equation}\label{baseMT_sum}
    m_i(t + \dt) = m_i(t) + \sum_{j = 1}^{N} \cW_{ij}\left[m_j(t) - m_i(t)\right],
\end{equation}
where $m_i(t)$ is the mass carried by particle $i$ at time $t$, $N$ is the number of particles, and
\begin{equation}\label{rho_norm}
    \cW_{ij} \defeq \frac{W(\vec X_i, \vec X_j; h)}{\rho_{ij}}.
\end{equation}
Note that this formulation is equivalent to choosing $\beta = 1$, in the context of \cite{guillem_SPH_equiv}.
Above, $W$ is a Gaussian weighting function that determines the amount of mass transferred from particle $j$ to particle $i$ (or vice-versa because $W$, in this case, is symmetric with respect to $\vec X_i$ and $\vec X_j$) and $\rho_{ij}$ is a normalizing constant that ensures conservation of mass.
We specify here that this normalization would not be required in the limiting, infinite-particle case, but for any finite number of particles, $N$ samples from the weighting function $W$ (which is necessarily a density) will not sum to unity and thus not conserve mass.
As such, we normalize our discretized density according to \eqref{rho_norm}.

In the case of isotropic diffusion, we have
\begin{equation}\label{iso_linear_gaussian}
    W(\vec X_i, \vec X_j; (D_i + D_j) \dt) = \left(2 \pi (D_i + D_j) \dt\right)^{-d/2} \exp\left[-\frac{\Vert \vec X_j - \vec X_i \Vert^2}{2 (D_i + D_j) \dt} \right],
\end{equation}
where $d = 1, 2, 3,$ is the number of spatial dimensions, and $D_k \defeq D(\vec X_k)$.
The matrix-vector form of \eqref{baseMT_sum} is written as
\begin{equation}\label{MT_arbT}
    \vec m(t + \dt) = \vec T \vec m(t),
\end{equation}
in which
\begin{equation}\label{Tmat}
    \vec T \defeq \vec I + \vec \cW - \diag(\vec \cW \vec 1),
\end{equation}
where $\vec I$ is the $N \times N$ identity matrix, $\vec 1$ is an $N \times 1$ vector of ones, and $\diag(\vec x)$ is a square matrix with the entries of vector $\vec x$ on its main diagonal.
A popular choice for $\rho_{ij}$ that results in symmetric $\vec \cW$ (and thus conservation of mass by the operator $\vec T$) is
\begin{equation}\label{symmNorm}
    \rho_{ij} \defeq \frac{[\vec W \vec 1]_{i} + [\vec 1^T \vec W]_j}{2},
\end{equation}
or, in words, $\rho_{ij}$ is the arithmetic mean of the sums of row $i$ and column $j$.

We see in the formulation outlined above that \eqref{iso_linear_gaussian} is the analytical solution to the diffusion equation over the interval $[0, \dt]$ as a function of position $\vec X_j$, given a unit point source located at $\vec X_i$ (or vice-versa, with respect to $\vec X_i$ and $\vec X_j$).
\rev{If we rewrite \eqref{iso_linear_gaussian} as
\begin{align}\label{iso_linear_gaussian_Deff}
    W(\vec X_i, \vec X_j; (D_i + D_j) \dt) &= \left(2 \pi (D_i + D_j) \dt\right)^{-d/2} \exp\left[-\frac{\Vert \vec X_j - \vec X_i \Vert^2}{2 (D_i + D_j) \dt} \right]\nonumber\\
    &= \left(4 \pi \left(\frac{D_i + D_j}{2}\right) \dt\right)^{-d/2} \exp\left[-\frac{\Vert \vec X_j - \vec X_i \Vert^2}{4 \left(\frac{D_i + D_j}{2}\right) \dt} \right]\nonumber\\
    &= \left(4 \pi \hat D \dt\right)^{-d/2} \exp\left[-\frac{\Vert \vec X_j - \vec X_i \Vert^2}{4 \hat D \dt} \right],
\end{align}
we see that his formulation computes an effective diffusion coefficient, $\hat D$, as the arithmetic mean of the diffusion coefficients at particle locations $\vec X_i$ and $\vec X_j$ (or, equivalently, linearly interpolates the diffusion coefficients between these two points and chooses the value located at the midpoint).}
However, as discussed in \cite{labolle_1996}, this linear interpolation fails in the case of discontinuous diffusion coefficients without the inclusion of some sort of reflection scheme to account for the infinite divergence in $D$ at the interface.
Put simply, this method only yields favorable results when $D(\vec x)$ can be reasonably approximated with a linear fit over distances on the order of $\ell \defeq \sqrt{(D_i + D_j) \dt}$, and clearly a linear approximation of an infinitely steep gradient will not suffice.
\rev{We note here that if we employ a harmonic mean to compute $\hat D$ in \eqref{iso_linear_gaussian_Deff} we can obtain low-error results in 1D and in certain 2D cases, but this approximation is not reliable in general.}
As such, it would seem that we need a more flexible functional form for our weighting function $W$, and the best possible choice would be the analytical solution to the diffusion equation that accounts for discontinuities in $D(\vec x)$.

\subsubsection{Analytical solution for mass-transfer weight function} 
\label{ssub:analytical_sol}

\begin{figure}[t]
    \centering
    \subfloat[]{\includegraphics[width=0.25\textwidth]{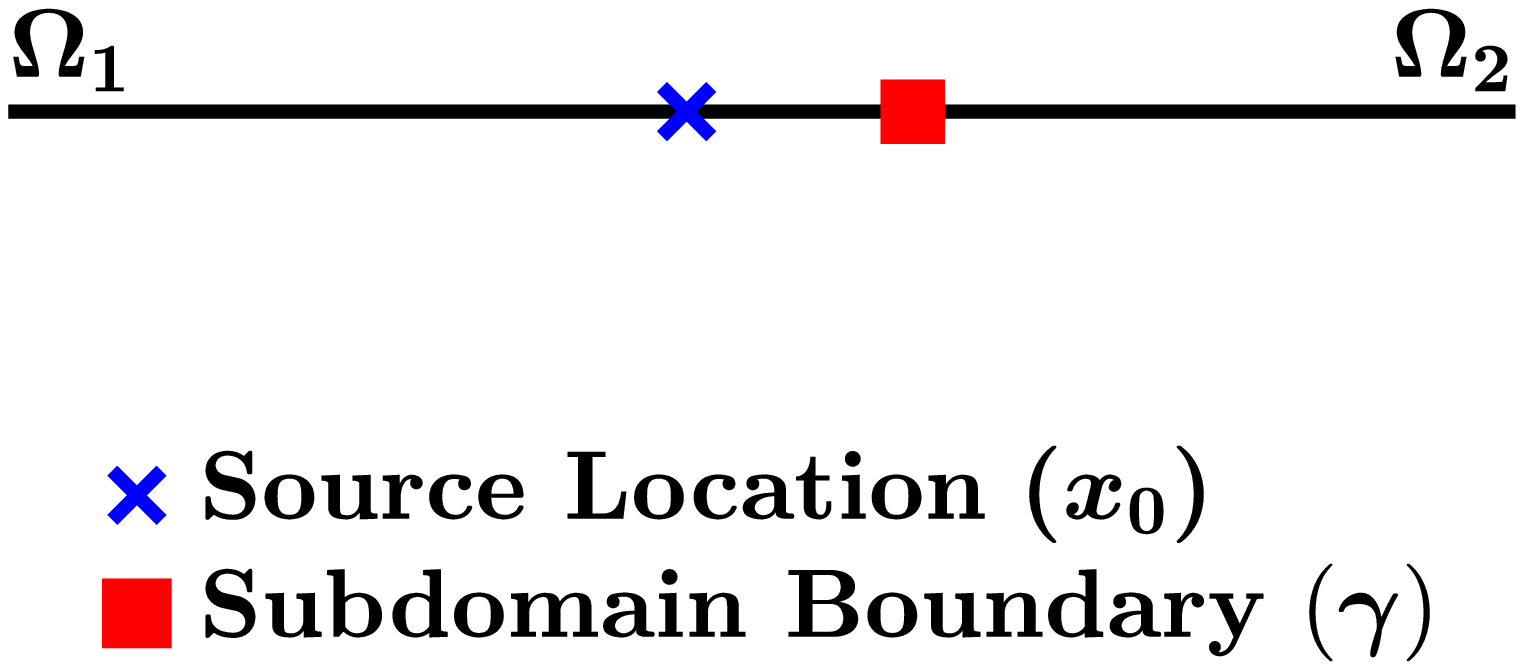}}
    \hspace{2.5em}
    \subfloat[]{\includegraphics[width=0.25\textwidth]{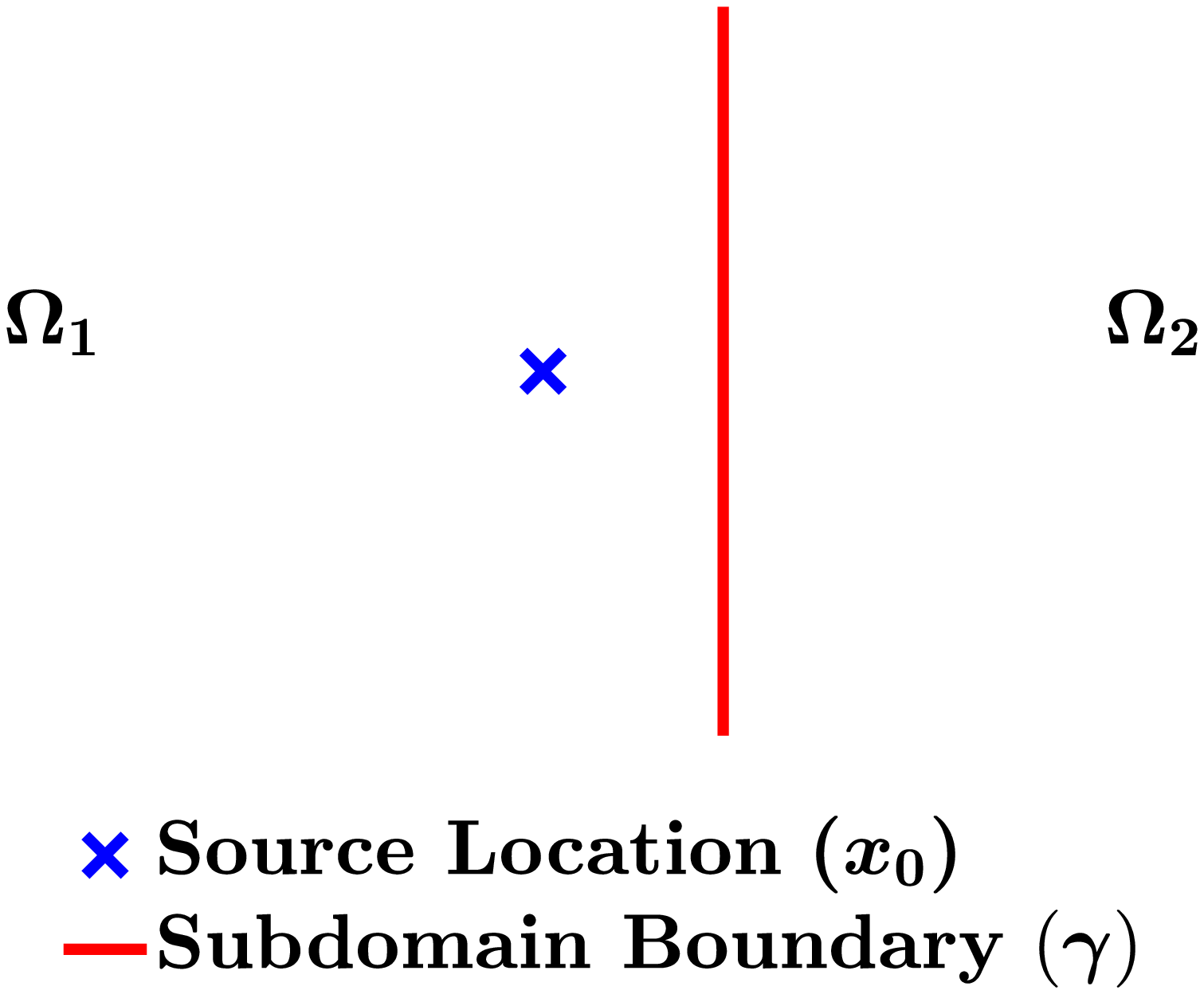}}
    \hspace{2.5em}
    \subfloat[]{\includegraphics[width=0.25\textwidth]{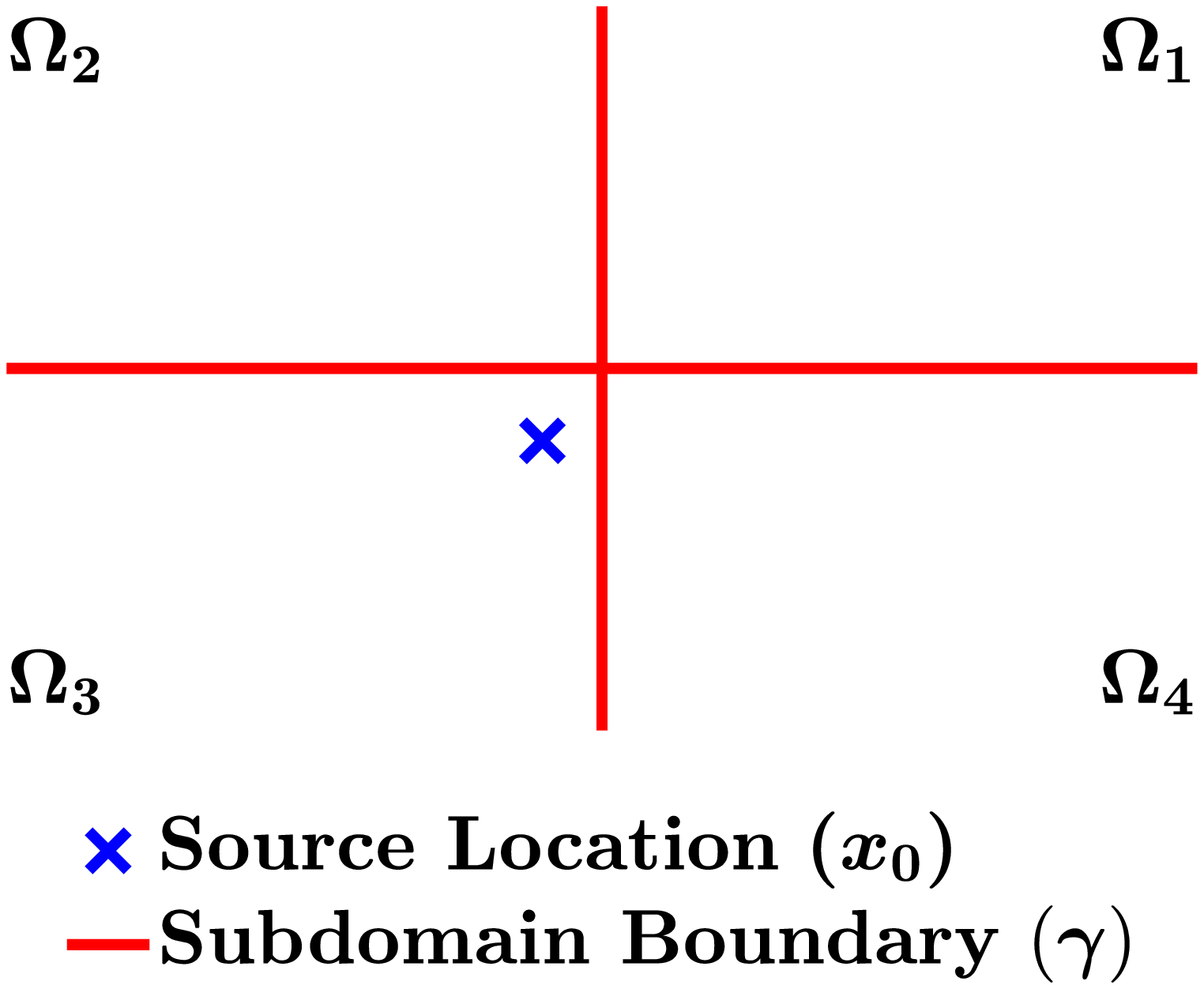}}
    \caption{Conceptual figures for the discontinuous diffusion coefficient problems we consider. (a) 1D problem with 2 subdomains, $\Omega_1$ and $\Omega_2$, with respective diffusion coefficients $D_1$ and $D_2$. The subdomains are split by the point $x = \gamma$, and the point-source initial condition is located at the point $x = \gamma$. (b) 2D problem with 2 subdomains, $\Omega_1$ and $\Omega_2$, with respective diffusion coefficients $D_1$ and $D_2$. The subdomains are split by the line $x = \gamma$, and the point-source initial condition is located at the point $\vec x = (x_0, y_0)$. (c) 2D problem with 4 subdomains, $\Omega_1, \Omega_2, \Omega_3, \Omega_4$, with respective diffusion coefficients $D_1, D_2, D_3, D_4$. The subdomains are split by the lines $x = \gamma_x$ and $y = \gamma_y$ and the point-source initial condition is located at the point $\vec x = (x_0, y_0)$.}
    \label{fig:cFigs}
\end{figure}

\emph{Carslaw and Jaeger} \cite{carslaw_jaeger} present a relatively simple solution in 1D for the problem of two subdomains.
We generalize that solution here for an instantaneous pulse of unit concentration at location $x = x_0 \in (-\infty, \infty)$ and time $t = 0$ (i.e., $C(t = 0, x) = \delta(x - x_0)$).
More specifically, for a chosen $\gamma \in (-\infty, \infty)$ we define the subdomains to be $\Omega_1 = (-\infty, \gamma]$ and $\Omega_2 = (\gamma, \infty)$, each with constant diffusion coefficients $D_1$ and $D_2$, respectively.
See Figure \ref{fig:cFigs}(a) for a conceptual depiction of this system.
If $x_0 \geq \gamma$, we have
\begin{equation}\label{CJsol_leftSource}
    C_A(t, x) = C_1(t, x; D_1, D_2)I_{\Omega_1}(x) + C_2(t, x; D_1, D_2)I_{\Omega_2}(x),
\end{equation}
where $I_A(z)$ is the indicator function on the set $A$, such that
\begin{equation}
    I_A(z) \defeq \begin{cases}
                        1,\quad z \in A \\
                        0,\quad z \notin A
                   \end{cases},
\end{equation}
and
\begin{align}
    C_1(t, x; D_1, D_2) &= \frac{D_2 D_1 (\pi D_1 D_2 t)^{-1/2}}{(D_2 \sqrt{D_1} + D_1 \sqrt{D_2})} \exp{\frac{-\left\vert x - \gamma - (x_0 - \gamma) \sqrt{D_1 / D_2} \right\vert^2}{4 D_1 t}},\label{CnJ_C1} \\
    C_2(t, x; D_1, D_2) &= \frac{1}{2\sqrt{\pi D_2 t}} \exp{\frac{-\vert x - x_0 \vert^2}{4 D_2 t}} + \phantom{x} \nonumber \\
                       &\phantom{=}\ \ \frac{D_2 \sqrt{D_1} - D_1 \sqrt{D_2}}{2 (D_2 \sqrt{D_1} + D_1 \sqrt{D_2}) \sqrt{\pi D_2 t}} \exp{\frac{-\vert x + x_0 - 2\gamma \vert^2}{4 D_2 t}},
\end{align}
and if $x_0 < \gamma$, the complementary solution is
\begin{equation}\label{CJsol_rightSource}
    \hat C_A(t, x) = C_2(t, x; D_2, D_1)I_{\Omega_1}(x) + C_1(t, x; D_2, D_1)I_{\Omega_2}(x).
\end{equation}
For the sake of compact notation, we may combine \eqref{CJsol_leftSource} and \eqref{CJsol_rightSource} into
\begin{equation}\label{CJsol_arbSource}
    W_A(t, x) \defeq \hat C_A(t, x; x_0) I_{\Omega_1}(x_0) + C_A(t, x; x_0) I_{\Omega_2}(x_0).
\end{equation}

We note that the solution given in \eqref{CJsol_arbSource} is not symmetric with respect to $x$ and $x_0$ (this is seen most clearly in the numerator of the exponential term in \eqref{CnJ_C1}); however, in application and due to the sharp decay in the exponential, \eqref{CJsol_arbSource} is typically symmetric to the order of machine precision.
As our objective is to eliminate errors, including those from a lack of symmetry, we alter the mass-transfer algorithm given in \eqref{baseMT_sum} such that
\begin{equation}\label{nonSymm_MT_sumForm}
    m_i(t + \dt) = m_i(t)+ \sum_{j = 1}^{N} \cW_{ij} m_j(t) - \sum_{j = 1}^{N} \cW_{ji} m_i(t),
\end{equation}
in which the mass of particle $i$ at time $t + \dt$ is its mass at time $t$ plus the sum of all the incoming mass-transfers, minus the sum of all outgoing mass-transfers.
Also, because $\cW_{ij} \neq \cW_{ji}$, we now strictly define $\cW_{ij}$ to be the normalized weight for the mass transfer from particle $j$ to particle $i$ (the converse is no longer true).
Equation \eqref{nonSymm_MT_sumForm} may be rewritten in an analogous form to \eqref{MT_arbT}, with
\begin{equation}\label{nonSymm_Tmat}
    \vec T \defeq \vec I + \vec \cW - \diag\left(\vec 1^T \vec \cW\right).
\end{equation}
If we use \eqref{CJsol_arbSource} as our weighting function in \eqref{nonSymm_Tmat}, again employing the symmetric normalization given in \eqref{symmNorm} to form $\vec \cW$ (because $\vec W$ is almost certainly symmetric to machine precision), then we obtain a mass-transfer method that generates very little error in simulating this system.
The algorithm for conducting a single mass transfer (within a timestep of length $\dt$) according to this method is given in Algorithm \ref{nonSymm_MTalg}, in which \texttt{WtFunction()} is defined to be \eqref{CJsol_arbSource}.

\begin{algorithm}[h]\caption{Mass-transfer Algorithm for Non-symmetric Weighting Function\label{nonSymm_MTalg}}
    \SetKwFunction{WtFunction}{WtFunction}
    \SetKwFunction{Sum}{Sum}
    \SetKwFunction{Diag}{Diag}
    \SetKwFunction{matMul}{matMul}
    \KwIn{Particle positions, $X = \vec X(t)$, and particle masses $m = \vec m(t)$.}
    \KwOut{Updated particle masses, $m = \vec m(t + \dt)$.}
    \DontPrintSemicolon \Comment*[r]{Build weight matrix}
    \For{i = 1 \KwTo N}{
        \For{j = 1 \KwTo N}{
            $W(i, j) =$ \WtFunction{$x_0 = X(j),\ x = X(i),\ \gamma,\ D_1,\ D_2,\ \dt$}
        }
    }
    \DontPrintSemicolon \Comment*[r]{Normalize weight matrix}
    \For{i = 1 \KwTo N}{
        \For{j = 1 \KwTo N}{
            $W(i, j) = W(i, j)\ /\ (\Sum(W(i, :))\ +\ \Sum(W(:, j))\ /\ 2)$
        }
    }
    \DontPrintSemicolon \Comment*[r]{Build transfer matrix}
    \For{i = 1 \KwTo N}{
        \For{j = 1 \KwTo N}{
            $T(i, j) = 1 + W(i, j) - \Sum(W(:, i))$ \Comment*[r]{$i^{\text{th}}$ column sum}
        }
    }
    $m = \matMul(T, m)$ \Comment*[r]{Conduct mass transfers}
\end{algorithm}

A major drawback of this method is that we must possess an analytic solution to the system of interest.
Granted, for small $\dt$, this solution is still relatively flexible; for example, we can still use this solution in the case of a 1D domain with three subdomains (considered in Section \ref{sec:results}), provided that \rev{the time step is sufficiently small or} the magnitude of diffusion in the center domain is low enough that mass-transfers do not ``see'' two subdomain boundaries at the same time.
Calculating an analytical solution is a non-trivial enterprise in spatial dimensions greater than one, particularly if we have a more complicated interface (for instance a $2 \times 2$ tiling of 4 subdomains in 2D, which we consider in Section \ref{sec:results}).
In fact, even for the relatively ``simple'' 2D problem of two half-planes, split by the line $x = \gamma$ (as considered in Section \ref{sub:2d_results}), the analytical solution is quite complex and likely infeasible as a mass-transfer kernel \cite[see][]{Shendeleva_discoD_2DSBS}.
As such, we seek a semi-analytical solution to the discontinuous $D(\vec x)$ problem, valid for small $\dt$, that will be flexible enough that it may be applied, by extension, to higher-dimensional problems.
We discuss this approach in the following section.


\subsubsection{Semi-analytical solution for mass-transfer weight function} 
\label{ssub:semi_analytical_sol}

In order to formulate our semi-analytical solution to the problem of a discontinuous diffusion coefficient, we take a predictor-corrector approach, much like that described in Section \ref{sub:RWPT_method} \cite{labolle_composite}.
We consider the same 1D problem setup outlined in Section \ref{ssub:analytical_sol}; however, for $x_0 \geq \gamma$, our semi-analytical solution shall have the form
\begin{equation}\label{semi_xGeqGamma}
    C_S(t, x) = C_r(x; D_1, \dt)I_{(-\infty, x_c]}(x) + C_k(x; D_2, \dt) I_{\Omega_2}(x),
\end{equation}
where the subscript $k$ stands for ``keep'' because this represents the amount of solute that is kept in the domain where it started (and is distributed according to a diffusion coefficient of $D_2$), and the subscript $r$ stands for ``redistribute'' because this represents the mass that is redistributed according to a diffusion coefficient of $D_1$, and $x_c$ is some ``corrected'' $x$-value that alters the support of the $C_r$ solution so that \eqref{semi_xGeqGamma} conserves mass.
Also, we make the distinction that $C_k$ and $C_r$ are parameterized by the necessarily small time step, $\dt$, rather than being functions of $t$, because this solution is only valid for short time.
In \eqref{semi_xGeqGamma}, we define
\begin{align}
    C_k(x; D_2, \dt) &\defeq \frac{1}{\sqrt{4 \pi D_2 \dt}} \exp\left[-\frac{\vert x - x_0 \vert^2}{4 D_2 \dt}\right],\\
    C_r(x; D_1, \dt) &\defeq \frac{1}{\sqrt{4 \pi D_1 \dt}} \exp\left[-\frac{\vert x - x_0 \vert^2}{4 D_1 \dt}\right].
\end{align}
Integrating each of these expressions over their respective support, in order to compute the total mass in each branch of the total solution, gives
\begin{align}
    m_k &= \int_{\gamma}^{\infty} C_k(x) dx = \frac{1}{2}\left[1 - \erf\left(\frac{\gamma - x_0}{\sqrt{4 D_2 \dt}}\right)\right], \\
    m_r &= \int_{-\infty}^{x_c} C_r(x)dx = \frac{1}{2}\left[1 - \erf\left(\frac{x_0 - x_c}{\sqrt{4 D_1 \dt}}\right)\right],
\end{align}
where $\erf(\cdot)$ is the error function.
Setting $m_{\text{total}} = 1 = m_k + m_r$ and solving for $x_c$ yields
\begin{equation}
    x_c = x_0 - (x_0 - \gamma) \sqrt{\frac{D_1}{D_2}},
\end{equation}
and we may repeat the calculations above for $x_0 < \gamma$, with the solution
\begin{equation}\label{semi_xLessGamma}
    \hat C_S(t, x) = C_k(x; D_1, \dt) I_{\Omega_1}(x) + C_r(x; D_2, \dt) I_{(x_c, \infty)}(x),
\end{equation}
to find
\begin{equation}
    x_c = x_0 - (x_0 - \gamma) \sqrt{\frac{D_2}{D_1}}.
\end{equation}
As in Section \ref{ssub:analytical_sol}, we may combine \eqref{semi_xGeqGamma} and \eqref{semi_xLessGamma} into one general solution, namely
\begin{equation}\label{semi_general}
    W_S(t, x) \defeq \hat C_S(\dt, x; x_0) I_{\Omega_1}(x_0) + C_S(\dt, x; x_0) I_{\Omega_2}(x_0).
\end{equation}

\begin{figure}[t]
    \centering
    \includegraphics[width = \textwidth]{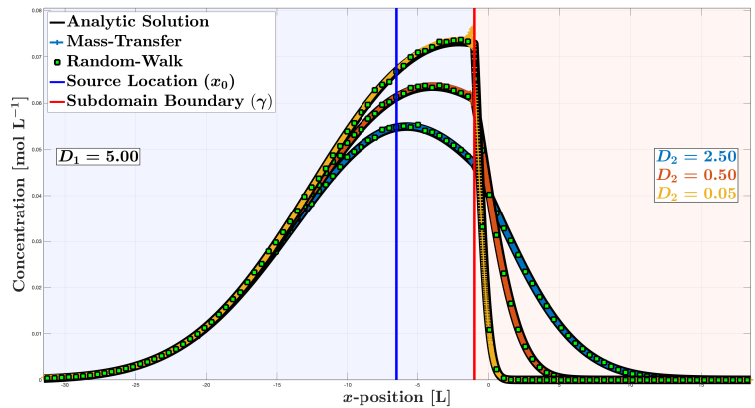}
    \caption{1D purely-diffusive simulation for two subdomains with diffusion coefficients $D_1$ and $D_2$ (shown for 3 different values of $D_2$).
    The MTPT method employs the semi-analytical solution given in \eqref{semi_general} using Algorithm \ref{nonSymm_MTalg}, and is compared to the predictor-corrector RWPT method of \cite{labolle_composite} and the analytical solution given in Section \ref{ssub:analytical_sol}.
    Results are shown for a simulation with $5000$ MT particles, $10^6$ RW particles, $\dt = 10^{-2}$, and total simulation time $T = 6$.
    All dimensioned quantities are unitless.
    Note the oscillation that occurs near the subdomain boundary ($x = \gamma$) for the $D_2 = 0.05$ case (yellow plot).
    This is attributable to applying Algorithm \eqref{nonSymm_MTalg} (symmetric normalization) rather than Algorithm \ref{mod_MTalg} (Sinkhorn-Knopp normalization).}
    \label{fig:1D_arbGamma_BAD}
\end{figure}

Unfortunately, if we use $W_S$ as the \texttt{WtFunction()} in Algorithm \eqref{nonSymm_MTalg}, we obtain solutions that display a troubling amount of oscillation near the subdomain boundary (see Figure \ref{fig:1D_arbGamma_BAD}).
This is because we no longer have a symmetric weight matrix (even numerically), due to $W_S$ lacking symmetry with respect to $x$ and $x_0$, and, as a result, it also no longer makes sense to apply the symmetric normalization given in \eqref{symmNorm}.
In order for the mass-transfer method to both conserve mass and generate solutions with low error, we must make the following changes:
\begin{enumerate}
    \item We normalize the weight matrix and form $\vec{\hat{\cW}}$ by employing the Sinkhorn-Knopp (SK) algorithm \cite{sink_knopp}, a computationally-efficient iterative method for obtaining a doubly-stochastic matrix that is mathematically equivalent to alternately normalizing the rows and the columns of a matrix to sum to unity.
    In order to conserve mass, the columns must be normalized last and must sum to unity with high precision.
    We find that for all of the cases we considered, 1000 iterations produced satisfactory results.
    \item We employ a weight matrix that is the transpose of that used in Algorithm \ref{nonSymm_MTalg}; i.e.,
    \begin{equation*}
        \hat{W}_{ij} \defeq W_S(\dt, x = X_j; x_0 = X_i).
    \end{equation*}
    To contrast, note that if we use \eqref{semi_general} in Algorithm \ref{nonSymm_MTalg}, we have
    \begin{equation*}
        W_{ij} \defeq W_S(\dt, x = X_i; x_0 = X_j).
    \end{equation*}
    This is done purely for numerical convenience, as applying the SK algorithm to $\vec{\hat{W}}$ converges more reliably to the desired stochastic matrix $\vec{\hat \cW}$ than applying SK to $\vec W$.
    In fact, starting with $\vec W$ leads to solutions that display a ``kink'' near the boundary, and much greater resolution in both time and space is required to generate acceptable solutions.
\end{enumerate}
Written in the sum form of \eqref{baseMT_sum} and \eqref{nonSymm_MT_sumForm}, after normalizing $\vec{\hat{W}}$ via SK to form $\vec{\hat{\cW}}$, the above amounts to
\begin{equation}\label{discoD_MT_sumForm}
    \begin{aligned}
        m_i(t + \dt) &= m_i(t)+ \sum_{j = 1}^{N} \hat{\cW}_{ij} m_j(t) - \sum_{j = 1}^{N} \hat{\cW}_{ji} m_i(t) \\
        &= m_i(t)+ \sum_{j = 1}^{N} \hat{\cW}_{ij} m_j(t) - m_i(t) \cancelto{1}{\sum_{j = 1}^{N} \hat{\cW}_{ji}} \\
        &= \sum_{j = 1}^{N} \hat{\cW}_{ij} m_j(t),
    \end{aligned}
\end{equation}
or in matrix-vector form we have
\begin{equation}
    \vec m(t + \dt) = \vec{\hat{\cW}} \vec m(t).
\end{equation}
The algorithm for conducting mass-transfers (within a timestep of length $\dt$), according to this modified method is presented in pseudocode in Algorithm \ref{mod_MTalg}, in which \texttt{WtFunction()} is defined to be \eqref{semi_general}.

\begin{algorithm}[t]\caption{Modified Mass-transfer Algorithm for Semi-analytical Weighting Function\label{mod_MTalg}}
    \SetKwFunction{WtFunction}{WtFunction}
    \SetKwFunction{colNormalize}{colNormalize}
    \SetKwFunction{rowNormalize}{rowNormalize}
    \SetKwFunction{Sum}{Sum}
    \SetKwFunction{Diag}{Diag}
    \SetKwFunction{matMul}{matMul}
    \KwIn{Particle positions, $X = \vec X(t)$, and particle masses $m = \vec m(t)$.}
    \KwOut{Updated particle masses, $m = \vec m(t + \dt)$.}
    \DontPrintSemicolon \Comment*[r]{Build weight matrix}
    \For{i = 1 \KwTo N}{
        \For{j = 1 \KwTo N}{
            $W(i, j) =$ \WtFunction{$x_0 = X(i),\ x = X(j),\ \gamma,\ D_1,\ D_2,\ \dt$}
        }
    }
    \DontPrintSemicolon \Comment*[r]{Normalize weight matrix}
    \For{i = 1 \KwTo normCount}{
        $W = \rowNormalize{W}$ \Comment*[r]{Normalize the rows of $W$}
        $W = \colNormalize{W}$ \Comment*[r]{Normalize the columns of $W$}
    }
    $m = \matMul(W, m)$ \Comment*[r]{Conduct mass transfers}
\end{algorithm}

We note that the normalization, conducted at lines 6-9 in Algorithm \ref{mod_MTalg}, is not strictly the SK algorithm, but is instead meant to be demonstrative, rather than computationally efficient.

\paragraph{Extension to 2D} 
\label{par:semi_sol_2D}

A major advantage of our semi-analytical solution is that it is  straightforward to extend to 2D by applying the same strategy as used in 1D.
Let us first consider the case of 2 subdomains that are split by the line $x = \gamma$, $\Omega_1 = \{(-\infty, \gamma] \times \R\}$ and $\Omega_2 = \{(\gamma, \infty) \times \R\}$ with respective constant diffusion coefficients $D_1, D_2$.
The initial condition is again the instantaneous point source $C(t = 0, \vec x) = \delta(\vec x - \vec x_0)$, and $\vec x_0 = (x_0, y_0)$.
See Figure \ref{fig:cFigs}(b) for a conceptual depiction of this system.
In this case, our general solution is nearly identical to the 1D case, namely
\begin{equation}\label{semiSol_2DSBS}
    W_S(t, \vec x) \defeq \hat C_S(\dt, \vec x; \vec x_0) I_{\Omega_1}(\vec x_0) + C_S(\dt, \vec x;  \vec x_0) I_{\Omega_2}(\vec x_0),
\end{equation}
where $C_S$ and $\hat C_S$ are the same as in \eqref{semi_xGeqGamma} and \eqref{semi_xLessGamma}, and the form of $C_k$ and $C_r$ are merely altered to contain 2D Gaussian functions; i.e.,
\begin{equation}\label{keep_redis_2D}
    C_i^{\text{2D}}(\vec x; \cD, \dt) \defeq \frac{1}{4 \pi \cD \dt} \exp\left[-\frac{\Vert \vec x - \vec x_0 \Vert^2}{4 \cD \dt}\right],\quad i = k, r,\quad \cD = D_1, D_2.
\end{equation}

The extension to a more complicated subdomain interface is also straightforward.
In this case, we consider a $2 \times 2$ tiling of 4 subdomains in 2D, and this condition captures the challenges presented by a highly heterogeneous diffusion (velocity) field that is discretized on a grid, perhaps generated by a finite-difference method.
Specifically, the challenge is that mass originating in a given quadrant can end up in any or all of the three neighboring quadrants, with the most complicated path being the diagonal one across the origin.
For this problem the full domain $\Omega$ is split along the lines $x = \gamma_x$ and $y = \gamma_y$.
Thus, we have $\Omega_1 = \{(\gamma_x, \infty] \times (\gamma_y, \infty)\},\ \Omega_2 = \{(-\infty, \gamma_x] \times (\gamma_y, \infty)\},\ \Omega_3 = \{(-\infty, \gamma_x] \times (-\infty, \gamma_y]\},\ \Omega_4 = \{(\gamma_x, \infty) \times (-\infty, \gamma_y]\}$ with respective constant diffusion coefficients $D_1, D_2, D_3, D_4$.
Once again, the initial condition has the form $C(t = 0, \vec x) = \delta(\vec x - \vec x_0)$, with $\vec x_0 = (x_0, y_0)$.
See Figure \ref{fig:cFigs}(c) for a conceptual depiction of this system.
The general solution may be written
\begin{equation}\label{semiSol_2x2}
    \begin{aligned}
        W_S(t, \vec x) \defeq &C_S^{1}(\dt, \vec x;  \vec x_0) I_{\Omega_1}(\vec x_0) + C_S^{2}(\dt, \vec x; \vec x_0) I_{\Omega_2}(\vec x_0) + \phantom{x} \\
        &C_S^{3}(\dt, \vec x;  \vec x_0) I_{\Omega_3}(\vec x_0) + C_S^{4}(\dt, \vec x; \vec x_0) I_{\Omega_4}(\vec x_0).
    \end{aligned}
\end{equation}
Above, the portion of the solution corresponding to $\vec x_0 \in \Omega_1$ is composed of the sum of four local solutions with the form of \eqref{keep_redis_2D}, namely
\begin{equation}
    \begin{aligned}
        C_S^{1}(t, \vec x) \defeq &C_k^{\text{2D}}(\vec x; D_1, \dt) I_{\Omega_1}(\vec x) + \phantom{x} \\
        &C_r^{\text{2D}}(\vec x; D_2, \dt) I_{(-\infty, x_c^{12}] \times (\gamma_y, \infty)}(\vec x) + \phantom{x}\\
        &C_r^{\text{2D}}(\vec x; D_3, \dt) I_{(-\infty, x_c^{13}) \times (-\infty, y_c^{13})}(\vec x) + \phantom{x} \\
        &C_r^{\text{2D}}(\vec x; D_4, \dt) I_{(\gamma_x, \infty) \times (-\infty, y_c^{14}]}(\vec x),
    \end{aligned}
\end{equation}
where $x_c^{ij}$ and $y_c^{ij}$ are the $x$ and/or $y$ corrections for the mass-transfers from subdomain $\Omega_i$ to $\Omega_j$ and are calculated so as to ensure conservation of mass.
Similar to the 1D problem, we have
\begin{equation}
    \begin{aligned}
        x_c^{12} &= x_0 - (x_0 - \gamma_x)\sqrt{\frac{D_2}{D_1}},\\
        x_c^{13} &= x_0 - (x_0 - \gamma_x)\sqrt{\frac{D_3}{D_1}},\\
        y_c^{13} &= y_0 - (y_0 - \gamma_y)\sqrt{\frac{D_3}{D_1}},\\
        y_c^{14} &= y_0 - (y_0 - \gamma_y)\sqrt{\frac{D_4}{D_1}},
    \end{aligned}
\end{equation}
and the calculations are analogous for the portions of \eqref{semiSol_2x2} corresponding to the other subdomains.

As in the 1D case, the solutions given in \eqref{semiSol_2DSBS} and \eqref{semiSol_2x2} do not conserve mass if they are used as the \texttt{WtFunction()} in Algorithm \ref{nonSymm_MTalg}, but they do conserve mass and generate minimal error if they are used in Algorithm \ref{mod_MTalg}.





\section{Results} 
\label{sec:results}

In this section, we consider the results of applying the MTPT algorithm described in Section \ref{sub:MTPT_method} to solve a series of increasingly-complicated test problems involving discontinuous $D(\vec x)$.
To do this we constrain our tests to only the mass-transfer (MT) portion of the MTPT algorithm (i.e., stationary particles that do not random-walk), and we compare the results of our simulation to known solutions.
In the simple 1D case of 2 subdomains, we compare our MTPT results to an analytical solution, and in all other cases, we use the established RWPT predictor-corrector method of \cite{labolle_composite} as our baseline for comparison.

\rev{We note that the idealized case of stationary, evenly-spaced particles we consider does not appear to bear much resemblance to an actual Lagrangian, or particle-tracking, method in which particles are stochastically positioned due to random walk diffusion.
However, even when particles random-walk, the MT algorithm is fully deterministic within each timestep, and, in fact, is conceptually a finite difference scheme with a stochastic stencil.
In previous work, the evenly-spaced, stationary condition is shown to bear more similarity to the random-walking particle case than it does to a randomly-spaced, stationary condition \cite{guillem_SPH_equiv}.
The reason for this is that when particles random-walk, they are ``on average'' equally-spaced at any given time; whereas, randomly-spaced, stationary particles inevitably contain persistent gaps between particles that degrade solution accuracy.
As a result, in order to isolate the performance of the MT algorithm and analyze its accuracy, we choose to simulate the algorithm on evenly-spaced, stationary particles.}

From an algorithmic standpoint, we generate the MTPT results according to Algorithm \ref{mod_MTalg}, and we use the appropriate semi-analytical solution as \texttt{WtFunction()}.
For the MTPT case, we model the initial condition by assigning the mass corresponding to unit mass to the particle located at $x_0$, and in the RWPT case, we place all particles at location $x_0$, each with mass $1/N$.
We then simulate a purely diffusive system with discontinuous $D(\vec x)$ up to final time $T$.
For MTPT, constructing the numerical solution at final time is as simple as plotting the concentration on each particle versus its position; however, in the case of RWPT, particles must first be binned to construct concentrations (equal length in 1D and equal-area squares in 2D), and the number of bins was chosen in each case so as to balance between low resolution and noisiness.
Lastly, for simplicity, all dimensioned quantities are unitless.

All numerical simulations were conducted in MATLAB, using a MacBook Pro with a 2.9 GHz Intel Core i5 processor and 8 GB of RAM.
The code used to generate the results in this section is available at \url{http://doi.org/10.5281/zenodo.3706926} \cite{discoD_repo}.

\subsection{1D Results} 
\label{sub:1d_results}

We begin with the simplest case of a 1D domain with two subdomains, as described in Sections \ref{ssub:analytical_sol} and \ref{ssub:semi_analytical_sol}, and we hold $D_1 = 5.0$ constant while we test 3 values of $D_2$ ranging from half the magnitude of $D_1$ to two orders of magnitude smaller.
In the simulations, we employ $5000$ particles for the MTPT simulations and 1 million particles in the RWPT simulations (grouped into 100 bins for plotting).
We choose a timestep of length $\dt = 10^{-2}$ with a total simulation time $T = 6$.

\begin{figure}[t]
    \centering
    \includegraphics[width = \textwidth]{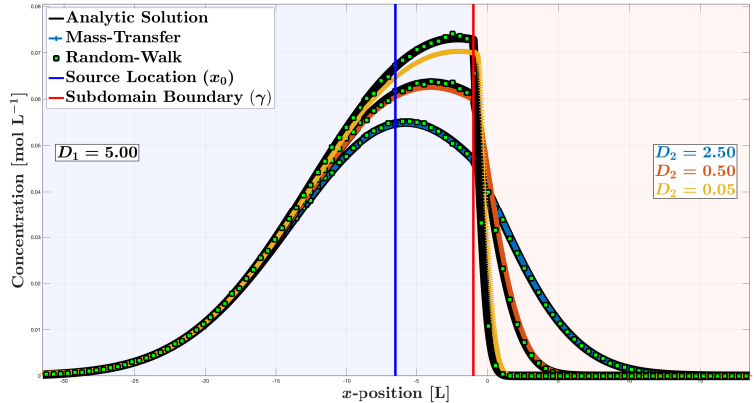}
    \caption{Results for a 1D purely-diffusive simulation for two subdomains with diffusion coefficients $D_1$ and $D_2$ (shown for 3 different values of $D_2$).
    The MTPT method employs the original MTPT algorithm on which we base our work \cite{Benson_arbitrary,mass_trans_acc,benson_entropy}, as compared to the predictor-corrector RWPT method of \cite{labolle_composite} and the analytical solution given in Section \ref{ssub:analytical_sol}.
    RW particles are grouped into 100 bins for plotting.
    Results are shown for a simulation with $5000$ MT particles, $10^6$ RW particles, $\dt = 10^{-2}$, and total simulation time $T = 6$.
    All dimensioned quantities are unitless.
    Note that the original MTPT algorithm performs quite poorly when there is a large disparity between $D_1$ and $D_2$.}
    \label{fig:1D_arbGamma_OG_MTPT}
\end{figure}

\begin{figure}[t]
    \centering
    \includegraphics[width = \textwidth]{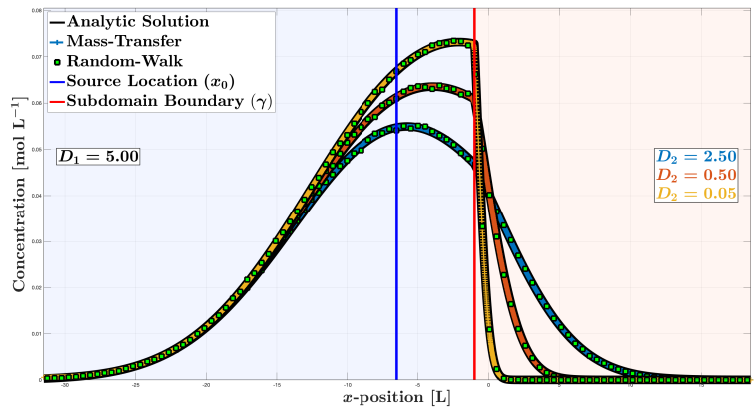}
    \caption{Results for a 1D purely-diffusive simulation for two subdomains with diffusion coefficients $D_1$ and $D_2$ (shown for 3 different values of $D_2$).
    The MTPT method employs the semi-analytical solution given in \eqref{semi_general} using Algorithm \ref{mod_MTalg}, as compared to the predictor-corrector RWPT method of \cite{labolle_composite} and the analytical solution given in Section \ref{ssub:analytical_sol}.
    RW particles are grouped into 100 bins for plotting.
    Results are shown for a simulation with $5000$ MT particles, $10^6$ RW particles, $\dt = 10^{-2}$, and total simulation time $T = 6$.
    All dimensioned quantities are unitless.}
    \label{fig:1D_arbGamma}
\end{figure}

We first examine what occurs when we apply the original MTPT method that our proposed algorithm is based upon (i.e., using \eqref{iso_linear_gaussian} as the weighting function in \eqref{Tmat}) \cite{Benson_arbitrary,mass_trans_acc,benson_entropy}.
These results are shown in Figure \ref{fig:1D_arbGamma_OG_MTPT}, and we see that the original MT algorithm holds up for a small magnitude difference in the diffusion coefficient, as when $(D_1, D_2) = (5, 2.5)$.
However, when the disparity becomes larger, the accuracy deteriorates, and the MTPT solution is quite poor for $(D_1, D_2) = (5, 0.05)$, as compared to the analytical solution and the RWPT results.
The results of applying our new MT algorithm (i.e., the semi-analytical solution given in \eqref{semi_general} used within Algorithm \eqref{mod_MTalg}) are depicted in Figure \ref{fig:1D_arbGamma}.
Comparing MTPT results both to the analytical solution, given in \eqref{CJsol_arbSource}, and the RWPT results, we see very close agreement between all solutions, indicating that our proposed approach is successful here and that we may move on to more complicated cases.

\begin{figure}[t]
    \centering
    \includegraphics[width = \textwidth]{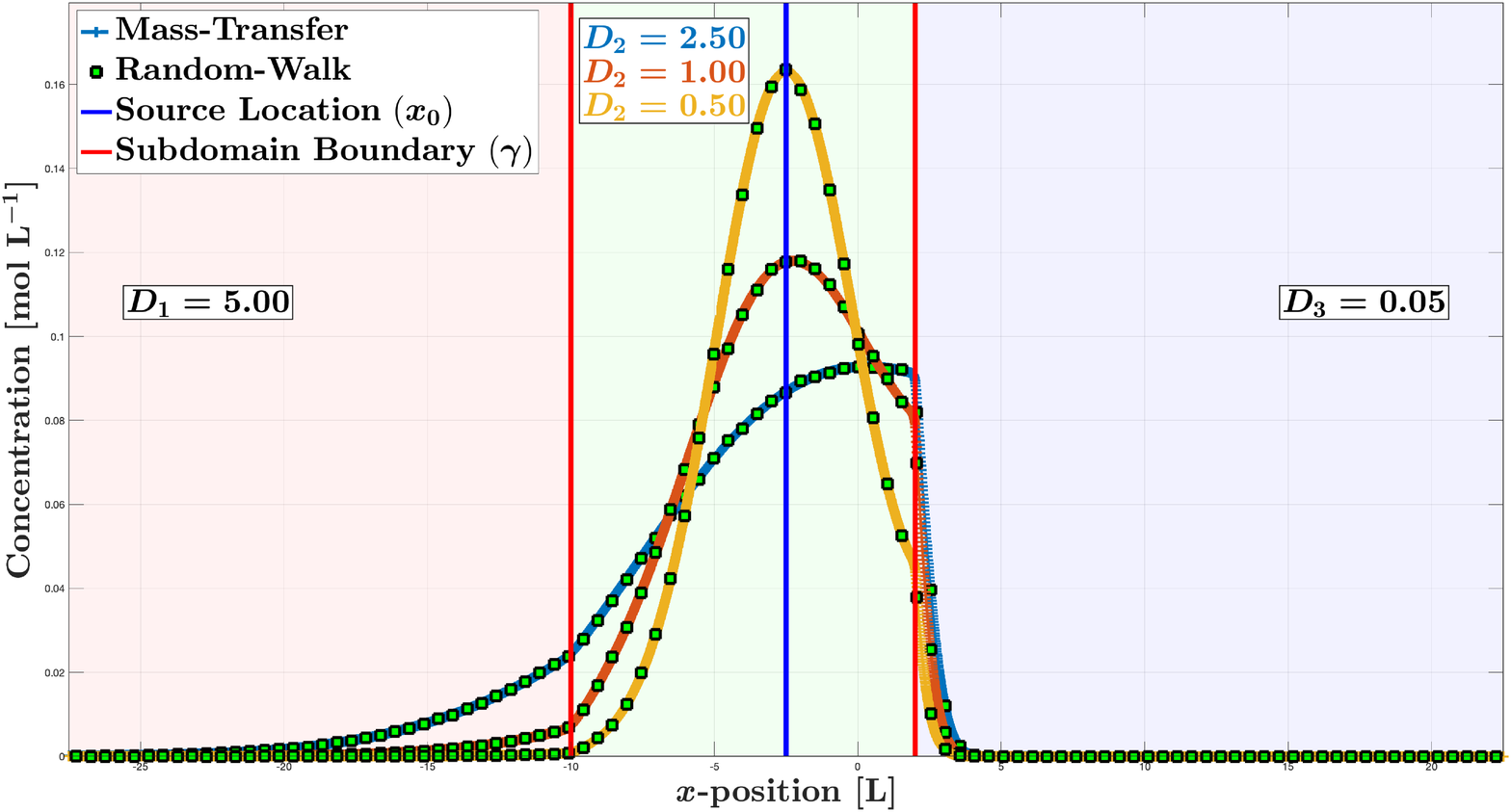}
    \caption{Results for a 1D purely-diffusive simulation for three subdomains with diffusion coefficients $D_1$, $D_2$, and $D_3$ (shown for 3 different values of $D_2$).
    The MTPT method employs the semi-analytical solution given in \eqref{semi_general} using Algorithm \ref{mod_MTalg}, as compared to the predictor-corrector RWPT method of \cite{labolle_composite}.
    RW particles are grouped into 100 bins for plotting.
    Results are shown for a simulation with $5000$ MT particles, $10^6$ RW particles, $\dt = 10^{-2}$, and total simulation time $T = 6$.
    All dimensioned quantities are unitless.}
    \label{fig:1D_triDomain}
\end{figure}

The next experiment we conduct focuses on a 1D problem with three subdomains, $\Omega_1$, $\Omega_2$, and $\Omega_3$, with their own respective diffusion coefficients, representing diffusion in, for example, a layered system.
We hold $D_1 = 5.0$ and $D_3 = 0.05$, so as to span two orders of magnitude, and we test three values of $D_2 \in \{2.5, 1.0, 0.5\}$ in the central subdomain.
In the simulations, we employ $5000$ particles for the MTPT simulations and 1 million particles in the RWPT simulations (grouped into 100 bins for plotting), and we choose a timestep of length $\dt = 10^{-2}$ with a total simulation time of $T = 6$.
The results of this experiment are displayed in Figure \ref{fig:1D_triDomain}.
Because this problem has no simple analytical solution, we take the RWPT results as our baseline case and find very close agreement of the MTPT results with the baseline.

\subsection{2D Results} 
\label{sub:2d_results}

\begin{figure}[t]
    \centering
    \includegraphics[width = \textwidth]{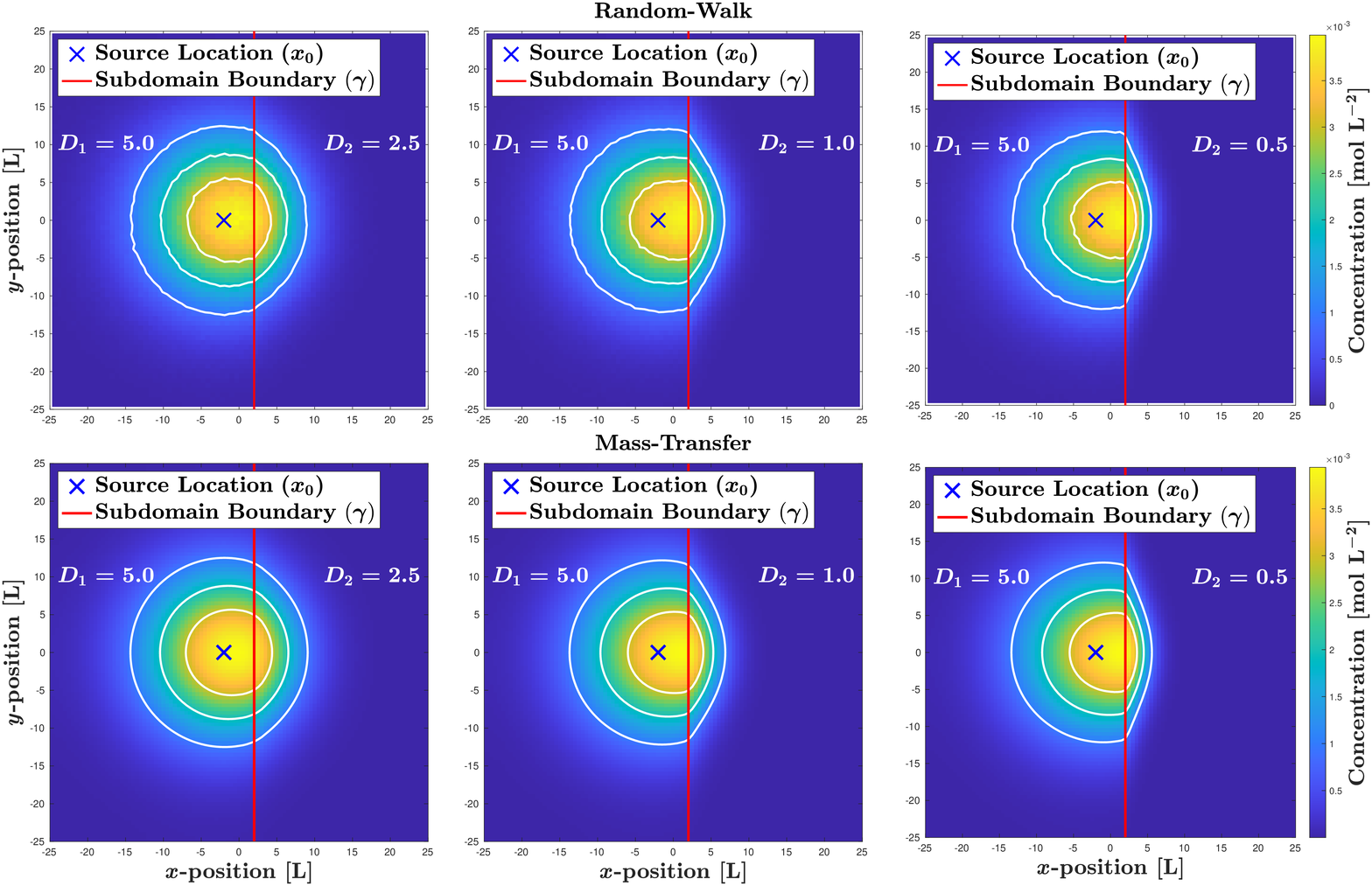}
    \caption{Concentration heatmap (magnitude given by the color bar on the righthand side) with constant-concentration contours (white curves) depicting results of a 2D simulation for two subdomains with diffusion coefficients $D_1$ and $D_2$ (shown for 3 different values of $D_2$).
    The MTPT method employs the semi-analytical solution given in \eqref{semiSol_2DSBS} using Algorithm \ref{mod_MTalg}, as compared to the predictor-corrector RWPT method of \cite{labolle_composite}.
    RW particles are grouped into 6400 bins for plotting.
    Results are shown for a simulation with $10201$ MT particles, $10^7$ RW particles, $\dt = 10^{-1}$, and total simulation time $T = 6$.
    All dimensioned quantities are unitless.}
    \label{fig:2D_SBS_heatMap}
\end{figure}

\begin{figure}[t]
    \centering
    \includegraphics[width = \textwidth]{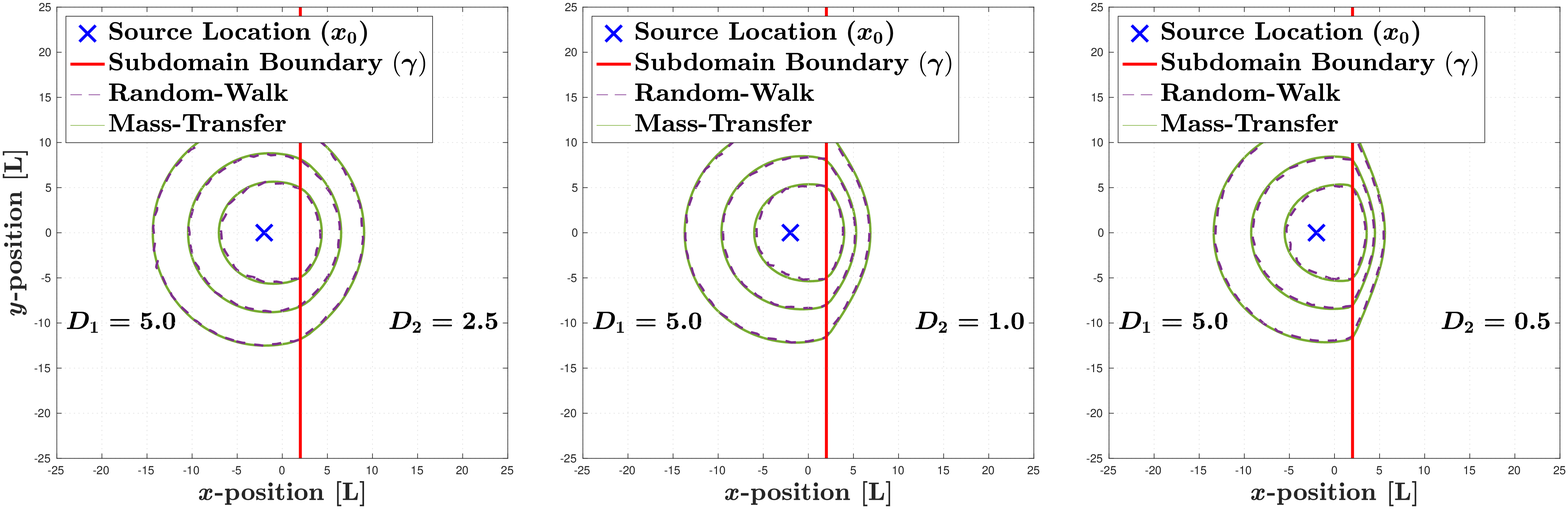}
    \caption{Constant-concentration contours comparing results depicting results of a 2D simulation for two subdomains with diffusion coefficients $D_1$ and $D_2$ (shown for 3 different values of $D_2$).
    The MTPT method employs the semi-analytical solution given in \eqref{semiSol_2DSBS} using Algorithm \ref{mod_MTalg}, as compared to the predictor-corrector RWPT method of \cite{labolle_composite}.
    RW particles are grouped into 6400 bins for plotting.
     Results are shown for a simulation with $10201$ MT particles, $10^7$ RW particles, $\dt = 10^{-1}$, and total simulation time $T = 6$.
     All dimensioned quantities are unitless.}
    \label{fig:2D_SBS_compare}
\end{figure}

Moving to 2D, we first consider the case of 2 subdomains split along the line $x = \gamma$, corresponding to the semi-analytical solution given in \eqref{semiSol_2DSBS}.
For these simulations, we hold $D_1 = 5.0$ and test $D_2 \in \{2.5, 1.0, 0.5\}$.
In the simulations, we employ $10201$ particles for the MTPT simulations (101 $\times$ 101 equally-spaced particles, with the number chosen so as to capture the integer-valued source location) and 10 million particles in the RWPT simulations (grouped into 6400 bins for plotting) and choose a timestep of length $\dt = 10^{-1}$ with a total simulation time of $T = 6$.
The results of this experiment are shown in Figures \ref{fig:2D_SBS_heatMap} and \ref{fig:2D_SBS_compare}.
In Figure \ref{fig:2D_SBS_heatMap}, we see good visual agreement of the MTPT solutions to the RWPT baseline, and this is verified by plotting the constant-concentration contours on the same axes in Figure \ref{fig:2D_SBS_compare} where the match is seen to be nearly exact, aside from the slight noise induced by the randomness in the RWPT simulation.

The next problem we consider is the 2D example of 4 subdomains split along the lines $x = \gamma_x$ and $y = \gamma_y$, corresponding to the semi-analytical solution given in \eqref{semiSol_2x2}.
For these simulations, the four cases we consider, in terms of choices for $D_i,\ i = 1, \dots, 4$, are: (1) 4 different values for $D_i$, spanning an order of magnitude; and 3 equal values for $D_i$ and one value that is an order of magnitude smaller, with (2) source location in a subdomain laterally adjacent to the small value of $D_i$, (3) source location in the subdomain containing the small value value of $D_i$, and (4) source location in a subdomain diagonally adjacent to the small value of $D_i$.
Of these four cases, case (4) is the least interesting, as the majority of solute remains in the three subdomains with large $D_i$, so we do not depict results of this simulation, though they were always favorable.
In the simulations, we employ $40401$ particles for the MTPT simulations (201 $\times$ 201 equally-spaced particles) and 10 million particles in the RWPT simulations (grouped into 6400 bins for plotting), and we choose a timestep of length $\dt = 10^{-1}$ for the MTPT simulations and $\dt = 10^{-2}$ for the RWPT simulations (this was required to generate smooth enough results for comparison), with a total simulation time of $T = 3$.

\begin{figure}[t]
    \centering
    \includegraphics[width = \textwidth]{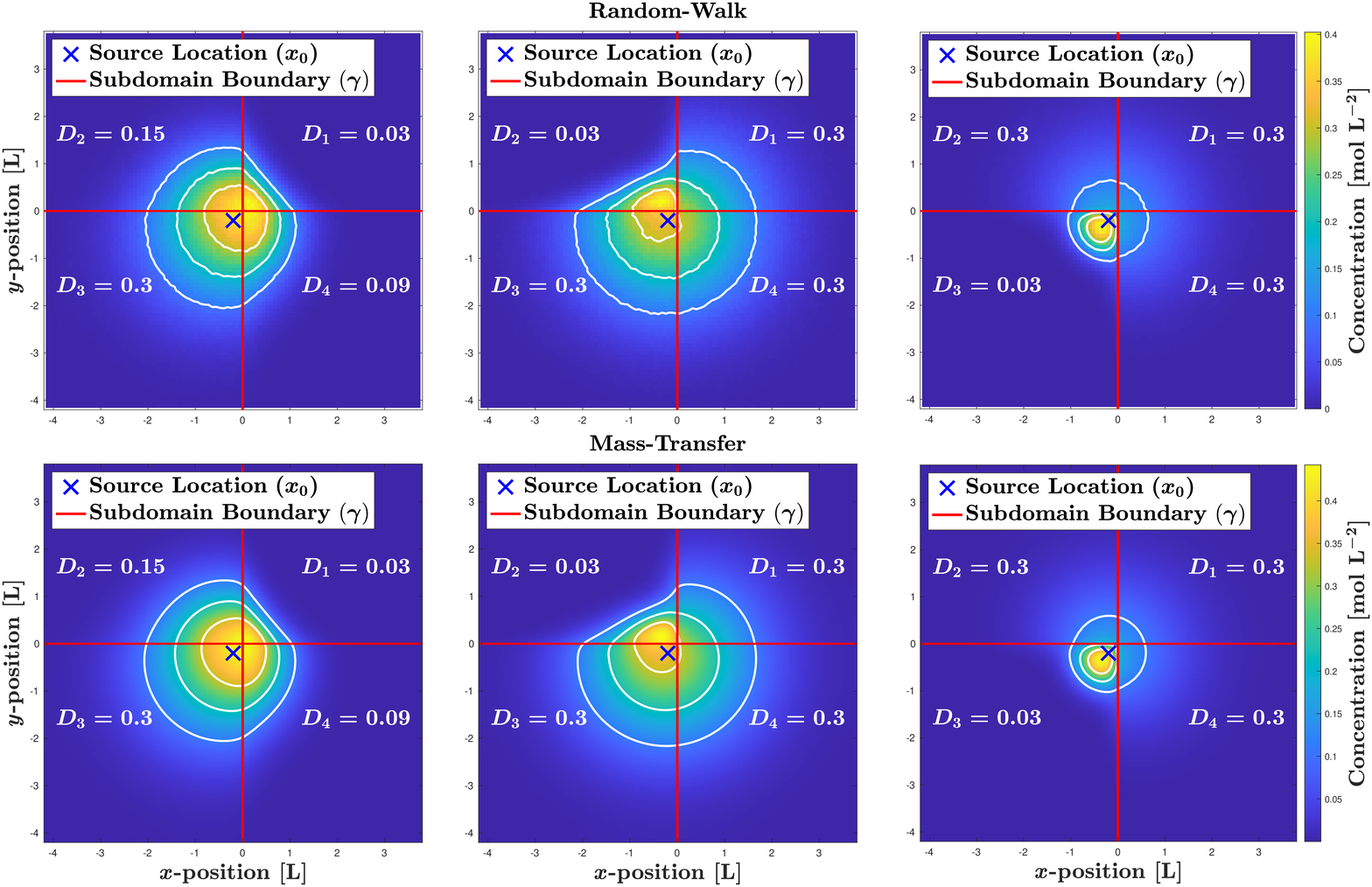}
    \caption{Concentration heatmap (magnitude given by the color bar on the righthand side) with constant-concentration contours (white curves) depicting results of a 2D simulation for four subdomains with diffusion coefficients $D_1$, $D_2$, $D_3$, and $D_4$.
    The MTPT method employs the semi-analytical solution given in \eqref{semiSol_2x2} using Algorithm \ref{mod_MTalg}, as compared to the predictor-corrector RWPT method of \cite{labolle_composite}.
    RW particles are grouped into 10201 bins for plotting.
    Results are shown for a simulation with $40401$ MT particles, $10^7$ RW particles, $\dt = 10^{-2}$, and total simulation time $T = 3$.
    All dimensioned quantities are unitless.}
    \label{fig:2D_2x2_heatMap}
\end{figure}
\begin{figure}[t]
    \centering
    \includegraphics[width = \textwidth]{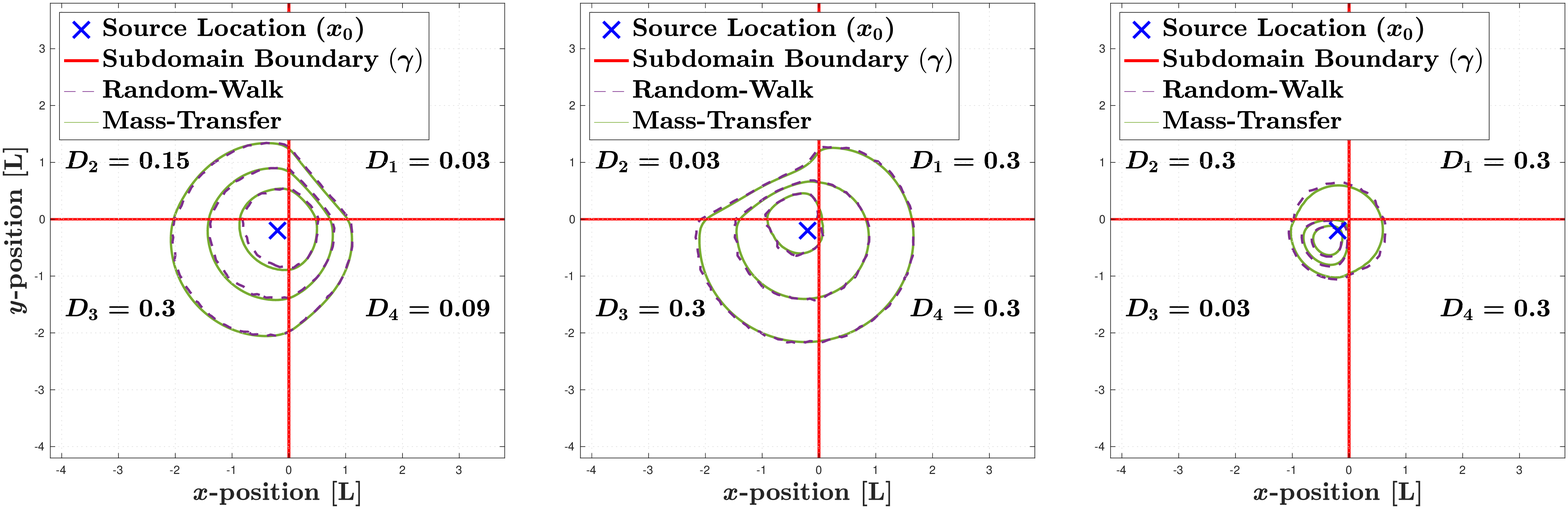}
    \caption{Constant-concentration contours comparing results depicting results of a 2D simulation for four subdomains with diffusion coefficients $D_1$, $D_2$, $D_3$, and $D_4$.
    The MTPT method employs the semi-analytical solution given in \eqref{semiSol_2x2} using Algorithm \ref{mod_MTalg}, as compared to the predictor-corrector RWPT method of \cite{labolle_composite}.
    RW particles are grouped into 10201 bins for plotting.
    Results are shown for a simulation with $40401$ MT particles, $10^7$ RW particles, $\dt = 10^{-2}$, and total simulation time $T = 3$.
    All dimensioned quantities are unitless.}
    \label{fig:2D_2x2_compare}
\end{figure}
The results of this experiment are shown in Figures \ref{fig:2D_2x2_heatMap} and \ref{fig:2D_2x2_compare}.
In Figure \ref{fig:2D_2x2_heatMap}, we see favorable visual agreement of the MTPT solutions to the RWPT baseline, and this is confirmed by the overlaid constant-concentration contour plots depicted in Figure \ref{fig:2D_2x2_compare}.
We note that in the 2D experiments, we only consider a single order or magnitude difference between diffusion coefficients.
This was in favor of fast run times, as the required number of particles for a MTPT simulation is dictated by the inter-particle spacing, which must be on the order of $\ell \defeq \sqrt{2 \tilde D \dt}$, where $\tilde D$ is the smallest diffusion coefficient in the system.
However, there are no theoretical barriers to considering larger disparities in $D(\vec x)$.


\subsection{Speed and Accuracy} 
\label{sub:speed_and_accuracy}

\rev{Here, we address two measures of algorithmic performance for our proposed MTPT method for discontinuous $D(\vec x)$.
First, as to speed, run times for the MTPT method are consistently lower than those for corresponding RWPT solutions.
For example, to generate the 1D, 2 subdomain results discussed in Section \ref{sub:1d_results} and depicted in Figure \ref{fig:1D_arbGamma}, the MTPT simulations run approximately 4.5 times faster than the RWPT simulations to which the solutions are compared.
For the 2D, 4 subdomain case, discussed in Section \ref{sub:2d_results} and depicted in Figures \ref{fig:2D_SBS_heatMap} and \ref{fig:2D_2x2_compare}, the MTPT simulations run approximately 1.5 times faster than the RWPT simulations.
This speedup for MTPT can primarily be attributed to the fact that mass-transfer interactions only occur among nearest-neighbors, and this allows for speedup via sparse linear algebra.
Note, however, that these run time comparisons are for reference only, as both algorithms can be optimized in various ways, and that was not the goal of this work.}

\begin{figure}[t]
    \centering
    \includegraphics[width = \textwidth]{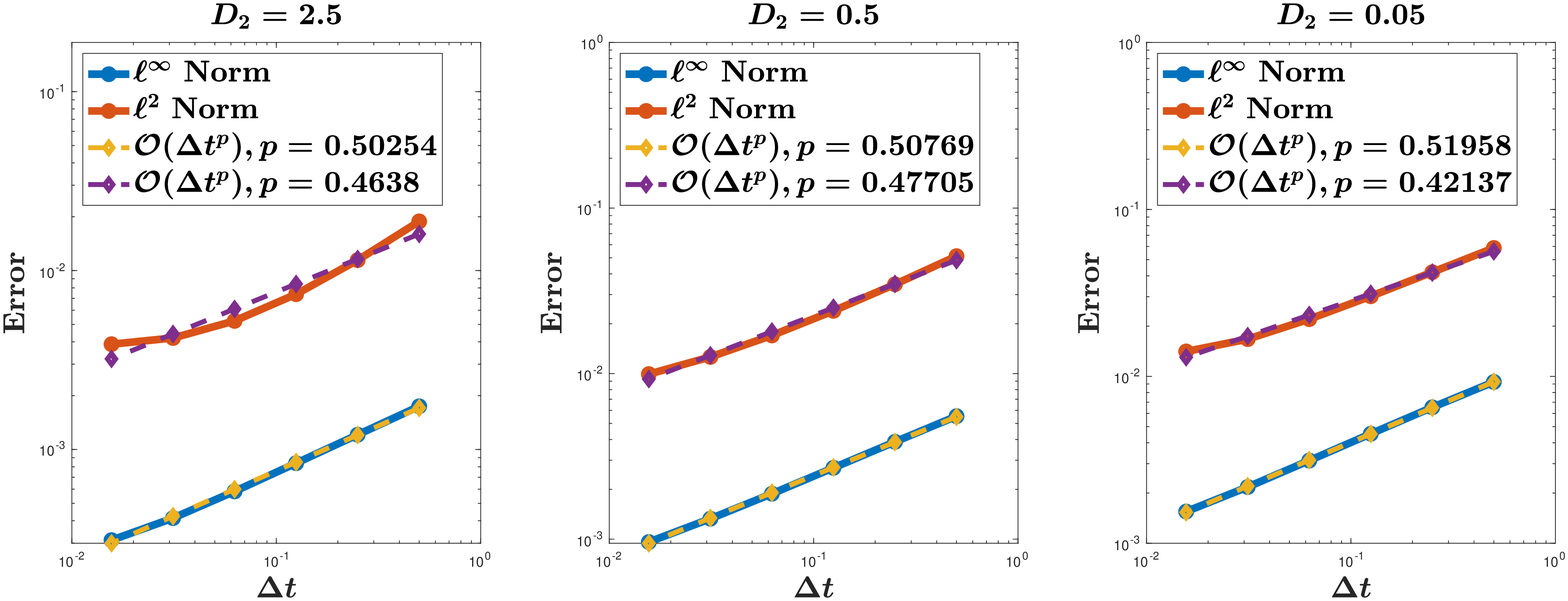}
    \caption{Convergence analysis in terms of time step length, $\dt$, for the MTPT Algorithm \ref{mod_MTalg} employing the semi-analytical solution given in \eqref{semiSol_2x2}. These results are for the 1D, 2 subdomain problem for which we have an analytical solution (see Section \ref{sub:1d_results} and Figure \ref{fig:1D_arbGamma}). Each plot corresponds to a single value for $D_2$. Error is computed in terms of the $\ell^2$ and $\ell^\infty$ norms and best-fit reference lines are shown to demonstrate the experimental order of convergence. $\cO(\dt^{1/2})$ convergence appears most clearly in the $\ell^\infty$ norm.}
    \label{fig:convDT}
\end{figure}

\begin{figure}[t]
    \centering
    \includegraphics[width = \textwidth]{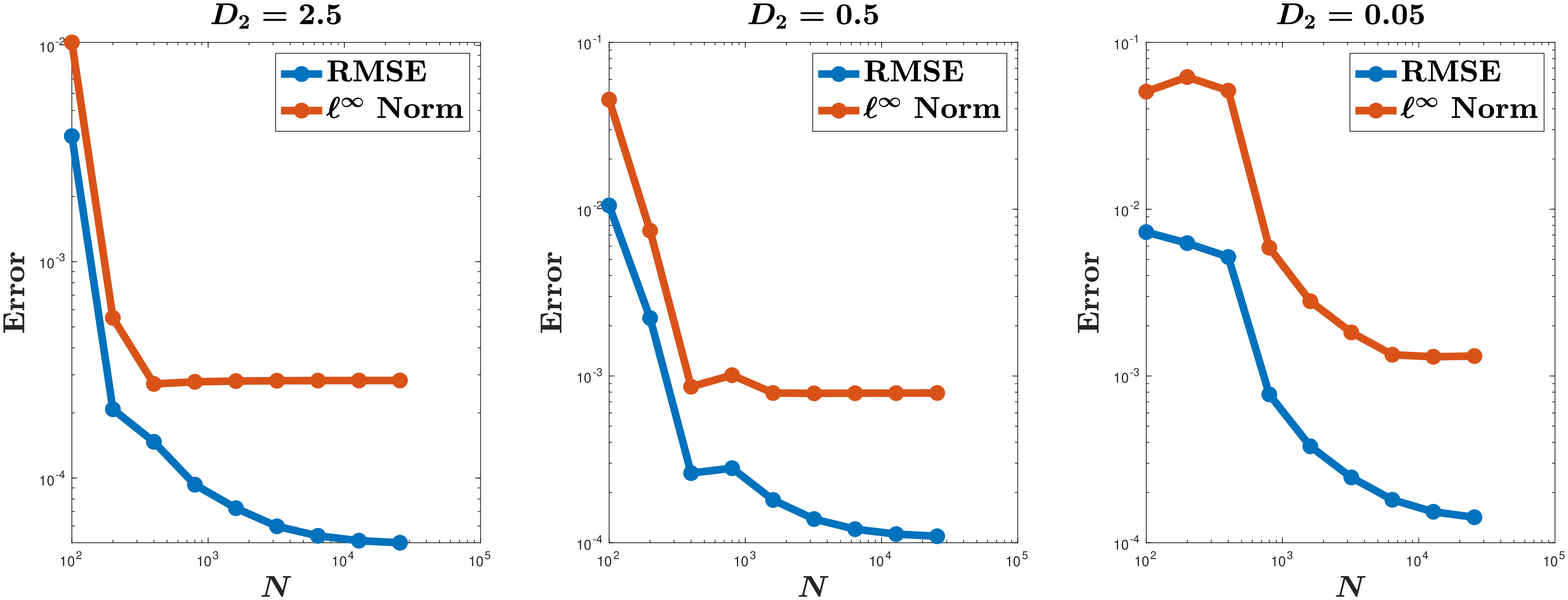}
    \caption{Convergence analysis in terms of particle number, $N$, for the MTPT Algorithm \ref{mod_MTalg} employing the semi-analytical solution given in \eqref{semiSol_2x2}. These results are for the 1D, 2 subdomain problem for which we have an analytical solution (see Section \ref{sub:1d_results} and Figure \ref{fig:1D_arbGamma}). Each plot corresponds to a single value for $D_2$. Error is computed in terms of the root-mean-squared error (RMSE) and $\ell^\infty$ norms. Depending on the value of $D_2$ being considered, rapid convergence is seen with increasing $N$, until leveling off at a minimal level.}
    \label{fig:convN}
\end{figure}

\rev{As to accuracy, we perform a convergence analysis for the proposed MTPT algorithm to see how error is affected by the level of discretization; i.e., refinements in time step length, $\dt$, or increase in particle number, $N$.
This convergence analysis considers the 1D, 2 subdomain case, and we compute error in comparison to the analytical solution given in \ref{ssub:analytical_sol}.
For each convergence analysis we employ all of the same parameters as were used to generate the results in Section \ref{sub:1d_results} and depicted in Figure \ref{fig:1D_arbGamma}, varying only $\dt$ or $N$ in successive refinements.
The results for a convergence analysis in terms of $\dt$ are depicted in Figure \ref{fig:convDT}, and therein we plot error, in terms of the $\ell^\infty$ and $\ell^2$ norms, as a function of $\dt$ for each of the three values of $D_2$ we consider in Section \ref{sub:1d_results} (Figure \ref{fig:1D_arbGamma}).
For each of the error curves we also plot a reference $\cO(\dt^p)$ line of best fit to obtain the order of convergence, $p$, and we see the general trend of what appears to be $p = 1/2$ order of convergence, and this is demonstrated most clearly in the $\ell^\infty$ norm.
The results for a convergence analysis in terms of $N$ are depicted in Figure \ref{fig:convN}, and therein we plot error, in terms of the $\ell^\infty$ norm and root-mean-squared error (RMSE), as a function of $N$ for each of the three values of $D_2$ we consider in Section \ref{sub:1d_results} (Figure \ref{fig:1D_arbGamma}).
We note that we employ RMSE here, as opposed to the $\ell^2$ norm, to normalize for varying vector-length.
In this case, we see a period of rapid convergence with increasing $N$, before error levels off to a minimal level that is controlled by the time discretization, and this is the expected behavior that is commonly seen in MTPT methods \cite{mass_trans_acc,Schmidt_fluid_solid}.
}

\section{Conclusions} 
\label{sec:conclusions}
Discontinuous diffusion coefficients arise naturally within simulations of transport through heterogeneous porous media, but accurately modeling diffusion across these interfaces has remained an outstanding problem for MTPT algorithms.
Here, we have generalized MTPT algorithms to addresses this deficiency, including for multi-dimensional systems.
This is a significant advance both from a numerical perspective and in terms of improving the realism of such simulations.
Additionally, these results serve to eliminate one of the few remaining barriers that limit the capabilities of Lagrangian methods in comparison to their Eulerian counterparts.

In particular, within the current work, we have:
\begin{enumerate}
    \item generalized the MT algorithm to incorporate non-symmetric mass-transfer kernels;
    \item presented an MT algorithm that employs a relatively simple 1D analytic solution to the discontinuous $D(x)$ problem;
    \item derived a semi-analytical solution to the discontinuous $D(\vec x)$ problem that is straightforward to generalize to higher dimensions and complicated subdomain interfaces;
    \item presented an MT algorithm that incorporates this semi-analytical solution;
    \item applied this updated MTPT algorithm to a variety of test cases, including a 2D problem that corresponds to a standard velocity grid with order-of-magnitude differences in $D(\vec x)$;
    \item attained favorable results of this application of the new MTPT algorithm.
\end{enumerate}
\rev{Additionally, while not considered in this work, it would be a simple matter to handle moving subdomain interfaces with this algorithm.
This is because particle interactions occur pairwise, and to make the relevant mass-transfer, the only required information is each particle's mass, position, and local diffusion coefficient, which are easy enough to establish within a timestep, no matter the current orientation of a subdomain boundary.}

Open questions remain in this direction, however.
For instance, what would be the effect of running a hybrid version of MTPT including diffusive random walks in the algorithm, and how would it affect the accuracy of solutions?
Or, how might the solution be generalized to subdomains that possess more complicated geometry; for example, boundaries that are not right angles, such as on a triangulated grid, or boundaries that are not straight lines at all \cite[e.g.][]{guillem_grid_project19}.
\rev{Additionally, we have only considered the scalar, or isotropic, $D(\vec x)$ case because it is common in the MTPT literature to simulate large-scale, anisotropic spreading by random walks and the micro-scale, isotropic mixing process by mass transfers \cite{mass_trans_acc,benson_mix_spread}.}

In summary, we have extended the capabilities of MTPT methods to solve the problem of discontinuous diffusion coefficients, thus adding flexibility to a tool that already is able to: model arbitrarily complex reactions, including fluid-solid interactions; separately simulate macro-scale spreading and micro-scale mixing; capture arbitrarily fine resolution in mixing and concentration gradients; and achieve nearly linear speedup when parallelized.


\section{Acknowledgments} 
\label{sec:acknowledgments}

We thank the editor and reviewers for their insightful and helpful comments.
The first author would like to thank Paul Martin for his assistance in exploring analytical solutions to the 2D problems considered in this work.

This work was supported by the US Army Research Office under contract/grant number W911NF-18-1-0338; the National Science Foundation under awards EAR-1417145 and DMS-1614586; and the DOE Office of Science under award DE-SC0019123.


\bibliography{discoD_bibl}

\end{document}